%% file: mguzzi-cipanp2018.tex
\documentclass[12pt]{article}
\usepackage{graphicx}
\usepackage{amssymb}

\newcommand\pubnumber{XXX-xxx-xx}
\newcommand\pubdate{\today}

\def\KSU{$^{(a)}$Department of Physics,\\ 
 Kennesaw State University, 370 Paulding Ave.,  30144 Kennesaw, GA, U.S.A.}
\def\support{\footnote{Presenter. CIPANP2018, Thirteenth Conference on the Intersections of Particle and Nuclear Physics, May 29 - June 3, 2018 Palm Springs, CA }}
\def\MSU{$^{(e)}$Department of Physics and Astronomy, \\ 
Michigan State University, East Lansing, MI 48824 U.S.A. }
\def\SMU{$^{(f)}$Department of Physics 
\\Southern Methodist University, Dallas, TX 75275-0181, U.S.A. }
\def\Taiwan{$^{(b)}$Department of Physics, 
\\National Tsing Hua University, Hsinchu City, Taiwan.}
\def\Urumqi{$^{(c)}$School of Physics Science and Technology \& Center for Theoretical Physics, \\ 
Xinjiang University, Urumqi, Xinjiang 830046 China. }
\def\INPAC{$^{(d)}$INPAC, Shanghai Key Laboratory for Particle Physics and Cosmology \\ \& School of Physics and Astronomy, Shanghai Jiao Tong University, Shanghai 200240, China.}
\def\Title#1{\begin{center} {\Large #1 } \end{center}}
\def\Author#1{\begin{center}{ \sc #1} \end{center}}
\def\Address#1{\begin{center}{ \it #1} \end{center}}

\newcommand\pubblock{\rightline{\begin{tabular}{l} \pubnumber\\
         \pubdate  \end{tabular}}}
\newenvironment{Abstract}{\begin{quotation}  }{\end{quotation}}

\def\Acknowledgements{\bigskip  \bigskip \begin{center} \begin{large}
             \bf ACKNOWLEDGEMENTS \end{large}\end{center}}

\input econfmacros.tex

\begin{document}
\begin{titlepage}
\pubblock

\vfill
\Title{\bf CTEQ-TEA parton distribution functions with intrinsic charm}
\vfill
\Author{Marco Guzzi$^{(a)}$\support, Tie-Jiun Hou$^{(b)}$, Sayipjamal Dulat$^{(c)}$, Jun Gao$^{(d)}$, Joey Huston$^{(e)}$, Pavel Nadolsky$^{(f)}$, Carl Schmidt$^{(e)}$, Jan Winter$^{(e)}$, Keping Xie$^{(f)}$, C.-P. Yuan$^{(e)}$}
\Address{\KSU}
\Address{\Taiwan}
 \Address{\Urumqi}
 \Address{\INPAC}
 \Address{\MSU}
 \Address{\SMU}

\vfill
\begin{Abstract}
\begin{center} 
{\bf Abstract} 
\end{center}

The possibility of a (sizable) nonperturbative contribution to 
the charm parton distribution function (PDF) in a nucleon is investigated 
together with theoretical issues arising in its interpretation. Results from the global PDF analysis are presented.
The separation of the universal component of the nonperturbative charm 
from the rest of the radiative contributions is discussed and 
the potential impact of a nonperturbative charm PDF on LHC scattering processes is illustrated. 
An estimate of nonperturbative charm magnitude in the CT14 and CT14HERA2 global QCD 
analyses at the next-to-next-to leading order (NNLO) in the QCD coupling strength 
is given by including the latest experimental data from HERA and the Large Hadron Collider (LHC).
A comparison between different models of intrinsic charm is shown and prospects for standard candle observables at the LHC are illustrated. 
\end{Abstract}
\vfill

\end{titlepage}
\def\thefootnote{\fnsymbol{footnote}}
\setcounter{footnote}{0}

\section{Introduction}

Global analyses of world experimental data use Quantum Chromodynamics (QCD) theory to analyze a wide
range of measurements, including precision data from HERA, the Tevatron, and the Large Hadron Collider (LHC).
Theory predictions for short-distance hard-scattering processes allow us to extract universal parton distribution functions (PDFs) of the proton within some approximations.
The determination of the parton content of the proton is challenged by the increasing precision of the recent LHC experimental measurements.
Precise proton PDFs from global QCD analyses are crucial for advanced tests of 
the Standard Model and to investigate possible physics beyond the Standard Model.
The most recent CTEQ-TEA (CT) global analysis of experimental data was recently published in Ref.~\cite{Dulat:2015mca} where the CT14NNLO PDFs (CT14 PDFs)
are determined from a perturbative QCD analysis at the next-to-next-to-leading order (NNLO) approximation. 

\subsection{The perturbative charm PDF}

Several QCD parameters such as $\alpha_s$ and the quark masses affect the global analysis   
and are correlated with the PDFs, which not only depend on the experimental data sets considered, but depend also
on the specific theory assumptions and underlying physics models.
As an example of one of these choices adopted in the CT analysis,
the charm quark and antiquark PDFs are assumed to be zero below the energy scale
$Q_c=Q_0$ that is of the order of the charm-quark mass $m_c$.
In the CT14 analysis, the charm quark and
antiquark PDFs are turned on at the scale $Q_c=Q_0=m_c=1.3$ GeV,
with an initial ${\cal O}(\alpha_s^2)$ distribution consistent with
NNLO matching~\cite{Buza:1995ie,Buza:1996wv} to the three-flavor
result. At higher energy scale $Q$, most of the charm PDF is generated from
the DGLAP evolution that proceeds through perturbative splittings of gluons and light-flavor quarks.
Therefore, in a standard global analysis the ``perturbative'' charm PDF is generated by 
perturbatively evolving the PDFs from the initial scale $Q_c$ to the experimental data scale $Q$.

\subsection{A nonperturbative component for the charm PDF}

Besides the perturbative charm PDF, the existence of power-suppressed (higher-twist) channels for charm quark production, that are independent of
the leading-power (twist-2, or perturbative) production of charm quarks, is rigorously predicted by the QCD theory.
An ``intrinsic charm'' (IC) quark component maybe generated by the nonperturbative structure of the hadronic bound state.
The dynamical origin of the IC and its magnitude have been extensively discussed in past and recent 
literature and have been the subject of a long-standing debate.  
The IC quarks have been associated with the excited
$|uudc\overline{c}\rangle$ Fock state of the proton wave
function~\cite{Brodsky:1980pb,Brodsky:1981se,Pumplin:2005yf,Chang:2011vx,Blumlein:2015qcn,Brodsky:2015fna}
and predicted by meson-baryon models~\cite{Navarra:1995rq,Paiva:1996dd,Steffens:1999hx,Hobbs:2013bia}.
The range of validity of the PDF models with nonperturbative charm has been studied in a recent CTEQ-TEA analysis (CT14IC)  published in Ref.~\cite{Hou:2017khm} and in other recent 
works~\cite{Jimenez-Delgado:2014zga,Jimenez-Delgado:2015tma,Lyonnet:2015dca,Ball:2016neh,Ball:2017nwa}.

\subsection{Fitted charm and nonperturbative charm parametrizations}

Starting from the factorization theorem for DIS cross sections with massive
fermions, that is a fundamental QCD result, one can draw a consequential distinction between 
the ``fitted'' charm PDF parametrization and the nonperturbative charm PDF. 
The fitted charm PDF accounts for the nonperturbative charm plus other (possibly not universal) higher ${\cal O}(\alpha_s)$ higher power suppressed terms.
Since the perturbative charm PDF component cancels near the threshold up to a higher order, 
the fitted charm component may approximate for a missing higher-order term or a power-suppressed nonperturbative component. 
The genuine nonperturbative charm PDF instead, is defined by the means of power counting of radiative contributions to DIS.  
Assuming that this additional nonperturbative charm component can be factorized like the perturbative charm
component, one is able to examine how it differs from the perturbative charm, 
and how it depends on theoretical inputs in a global QCD analysis of PDFs.
In principle, the intrinsic charm content would be suppressed by powers of $(\Lambda^2_{\rm QCD}/m^2_c)$, but, since
this ratio is not very small, it may be relevant in some processes such as precise DIS.
For a more detailed description of QCD factorization with power suppressed charm contributions, 
we refer the reader to the recent CT14IC analysis of Ref.~\cite{Hou:2017khm}.

\subsection{Valence-like and sea-like models for the charm PDF}

According to the estimate of various models, the power-suppressed charm cross section is of the order of a fraction of the $\alpha_{s}^{2}$ component in DIS charm production, 
carrying less than about a percent of the proton's momentum.
The CT14IC analysis examines a more extensive list of nonperturbative models and it includes the most complete
set of DIS data from HERA as well as data from the LHC and (optionally) the EMC experiment~\cite{Aubert:1982tt} in the PDF fit.
Moreover, it utilizes a PDF parametrization that results in a more physical behavior of the PDFs.
Given that several mechanisms may give rise to the fitted charm, in the CT14IC analysis it is parametrized by two generic shapes,
a ``\emph{valence-like}'' and a ``\emph{sea-like}'' shape. These two shapes arise in a variety of dynamical models.

A valence-like shape has a local maximum at $x$ above
0.1 and satisfies $f_{q/p}(x,Q_{c})\sim x^{-a_1}$ 
with $a_1 \lesssim 1/2$
for $x\rightarrow0$ and $f_{q/p}(x,Q_{c})\sim (1-x)^{a_2}$ with $a_2 \gtrsim 3$
for $x\rightarrow 1$. The distributions for valence $u$ and $d$ quarks
fall into this broad category, as well as the ``intrinsic'' sea-quark
distributions that can naturally be generated in several ways
\cite{Pumplin:2005yf}, e.g.,
for all flavors, nonperturbatively from a $|uudQ\overline{Q}\rangle$
Fock state in light-cone
\cite{Brodsky:1980pb,Brodsky:1981se,Chang:2011vx,Blumlein:2015qcn,Brodsky:2015fna}
and meson-baryon models
\cite{Navarra:1995rq,Paiva:1996dd,Steffens:1999hx,Hobbs:2013bia}; for $\bar{u}$ and $\bar{d}$, from connected diagrams in lattice
QCD \cite{Liu:2012ch}. In contrast to the light flavors, in lattice QCD a charm PDF arises exclusively from disconnected diagrams \cite{KFLiuPrivate}. 
This suggests that $c$ and $\bar c$ contributions in DIS are connected to the hadron target by gluon insertions.
The approximate Brodsky-Hoyer-Peterson-Sakai (BHPS) model \cite{Brodsky:1980pb,Brodsky:1981se} parametrizes the charm PDF at $Q_{0}$ by a valence-like nonperturbative
function
\begin{equation}
\widehat{c}(x)=\frac{1}{2}A~x^{2}\left[\frac{1}{3}(1-x)(1+10x+x^{2})-2x(1+x)\ln{\left(1/x\right)}\right].\label{eq:modelB}
\end{equation}
This function is obtained from a light-cone momentum distribution by
taking the charm mass to be much heavier than the masses of the proton
and light quarks: $m_{c}\gg M_{p},m_{u},m_{d}$.  $A$ is the normalization factor that is to be determined from the fit.
The BHPS1 and BHPS2 global fits correspond to two different values of $A$ and are obtained with the parametrization choice of Eq.\ref{eq:modelB}. 
They are illustrated in Sec. \ref{results} below. 
Instead of approximating the probability integral as in the original BHPS model,
the $\widehat{c}(x)$ can also be obtained by solving the BHPS model
for the $|uudc\bar{c}\rangle$ Fock state numerically by keeping
the exact dependence on $M_{p},m_{u}$, and $m_{d}$.
In this BHPS model, the intrinsic quark distributions are 
determined by starting from a $|uudq\bar{q}\rangle$
proton Fock state, where the probability differential for a quark
$i$ to carry a momentum fraction $x_{i}$ is given by
\begin{equation}
d{\cal P}(x_{1},\dots,x_{5})=A\ dx_{1}\dots dx_{5}\
\delta(1-\sum_{i=1}^{5}x_{i})\frac{1}{\left[M_{p}^{2}-\sum_{i=1}^{5}\frac{m_{i}^{2}}{x_{i}^{2}}\right]^{2}}\,.\label{bhps-prob}
\end{equation}
This generalized BHPS model, used in the context of the CT14HERA fit with IC, is named BHPS3. 
However, the intrinsic contribution to the $s$ quark PDF is not included because
it is overwhelmed by the very large strange PDF uncertainty.
The presence of an intrinsic component for the strange quark does not affect our
conclusions about the nonperturbative charm.

A sea-like component is usually monotonic in $x$
and satisfies $f_{q/p}(x,Q_{c})\sim x^{-a_1}$ for $x\rightarrow0$
and $f_{q/p}(x,Q_{c})\sim (1-x)^{a_2}$ for $x\rightarrow 1$,
with $a_1$ slightly above 1, and $a_2 \gtrsim 5$. This behavior is typical for the
leading-power, or ``extrinsic'' production.
For example, an (anti)quark PDF with this
behavior originates from $g\rightarrow q\bar{q}$ splittings in perturbative
QCD, or from disconnected diagrams in lattice QCD (see Ref.~\cite{Liu:2012ch}
for details). Even a missing next-to-next-to-next-to-leading order (N3LO)
leading-power correction may produce a sea-like contribution at $x\ll 0.1$, where the valence-like
components are suppressed.
In the SEA model, the charm PDF is parametrized by a ``sea-like'' nonperturbative function that is proportional
to the light quark distributions:
\begin{equation}
\widehat{c}(x)=A~\left(\overline{d}(x,Q_{0})+\overline{u}(x,Q_{0})\right)\,.\label{eq:modelS}
\end{equation}
This model is assumed with the SEA1 and SEA2 PDF sets which are discussed in Sec. \ref{results}.
Finally, the normalization coefficient $A$ in the models described above can be derived from the charm momentum
fraction (first moment) at scale $Q$:
\begin{equation}
\langle{x}\rangle_{{\rm IC}}=\int_{0}^{1}x\left[c(x,Q_{0})+\bar{c}(x,Q_{0})\right]dx.\label{xICdef}
\end{equation}

By its definition, $\langle x\rangle_{\rm IC}$ is evaluated
at the initial scale $Q_{0}$ and it is to be distinguished from the full
charm momentum fraction $\langle x\rangle_{c+\bar{c}}(Q)$ at $Q > Q_c$,
which rapidly increases with $Q$ because of the combination with the twist-2 charm component.

\section{Results of the global analysis\label{results}}

In this section the main findings of the CT14IC global analysis are illustrated. Here it is assumed 
that the additional nonperturbative charm component can be factorized in a similar fashion to the perturbative one. 
\begin{figure}[tb]
\begin{center}
\includegraphics[width=0.45\textwidth]{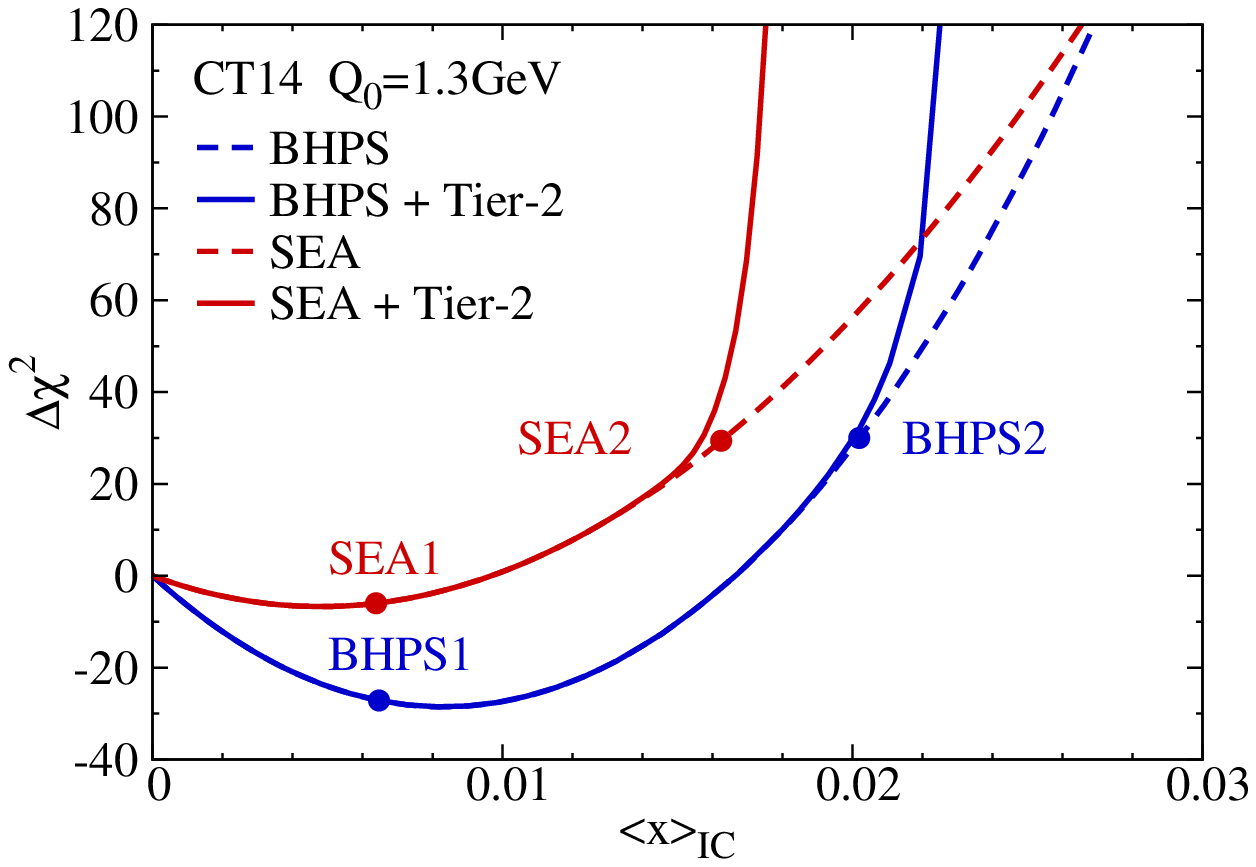}
\includegraphics[width=0.45\textwidth]{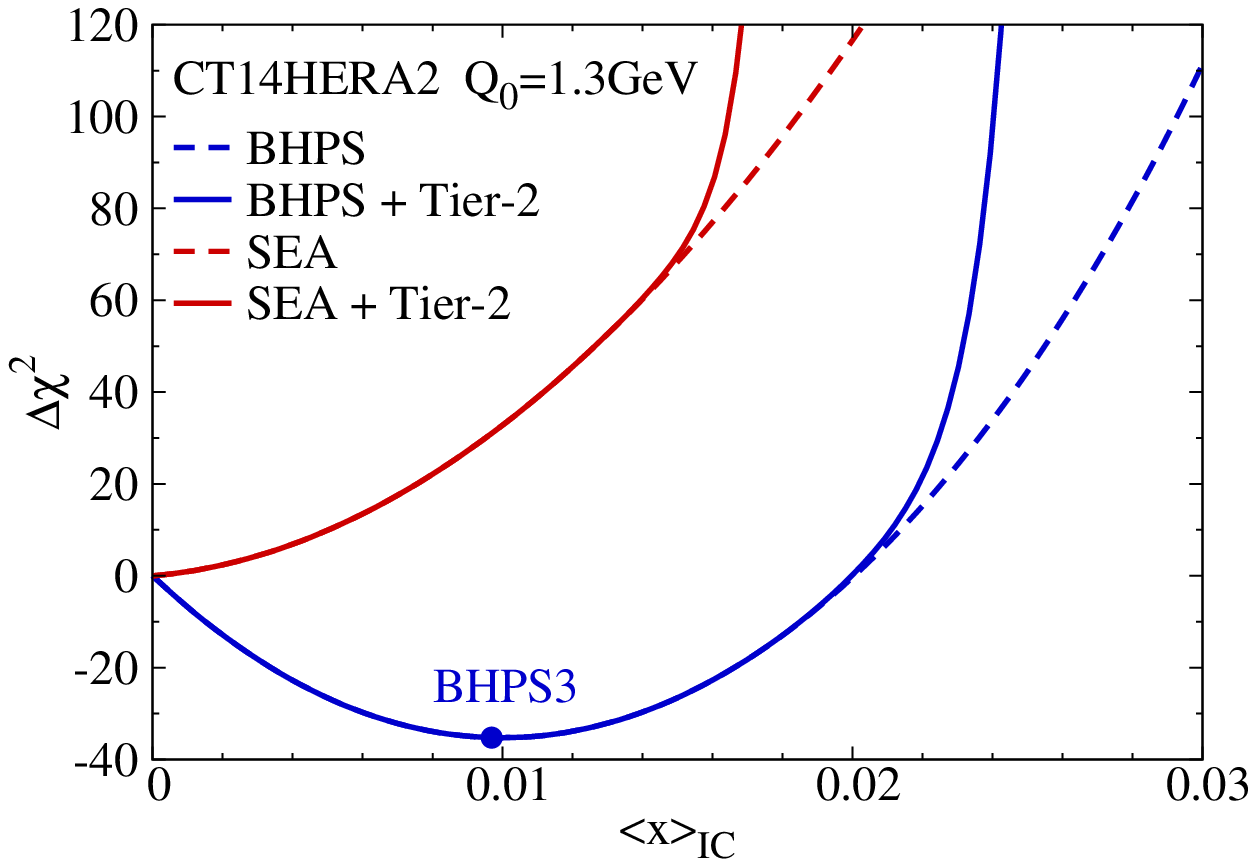}
\end{center}
\caption{The change $\Delta\chi^{2}$ in the goodness of fit to the
  CT14 (left) and CT14HERA2 (right) data sets
  as a function of the charm momentum fraction
  $\langle{x}\rangle_{\rm IC}$ for the BHPS (blue) and SEA (red) models.
Solid (dashed) lines represent the total $\chi^2$ and the partial
$\chi^2_{global}$, as defined in Sec.~\ref{results}.
\label{fig:delta_chisqVxic}}
\end{figure}
The goodness-of-fit function is defined as $\chi^2 \equiv \chi^2_{global} + P \label{chi2chi2P}$ 
and it is constructed from the global $\chi^2_{global}$ and a ``tier-2'' statistical penalty $P$ according to the CT14 method~\cite{Dulat:2015mca}.
This is used to estimate the preference of the global QCD data to a specific $\langle x\rangle_{\rm IC}$.
A convenient strategy is to compare each fit with an
$\langle x\rangle_{\rm IC}\neq 0$ to the ``null-hypothesis'' fit
obtained assuming $\langle x\rangle_{\rm IC}=0$. Thus, one starts by computing 
$\Delta \chi^2 \equiv \chi^2 - \chi^2_0, \label{Deltachi2}$
 where $\chi^2$ and $\chi^2_0$ are given for $\langle x\rangle_{\rm
  IC}\neq 0$ and $\langle x\rangle_{\rm
  IC}=0$, respectively, at 50 values of $\langle x\rangle_{\rm IC}$ and using the default
$Q_0=m_c^{pole}=1.3$ GeV. 
The resulting $\Delta \chi^2$ behavior is shown in
Fig.~\ref{fig:delta_chisqVxic}. The CT14
(CT14HERA2) data sets are compared against the
approximate and exact solution of the BHPS model.
The SEA charm parametrizations are constructed as in
Eq.~(\ref{eq:modelS}) in terms of the respective CT14 or CT14HERA2
light-antiquark parametrizations.
Fig.~\ref{fig:delta_chisqVxic} shows that large amounts
of intrinsic charm are disfavored for all models considered in the analysis.
A mild reduction in $\chi^{2}$, however, is observed for the BHPS
fits, roughly at $\langle{x}\rangle_{\rm IC} = 1\%$,
both in the CT14 and CT14HERA2 frameworks.
The significance of this reduction and the upper
limit on $\langle x\rangle_{\rm IC}$
depends on the assumed criterion. In CTEQ
practice, a set of PDFs with $\Delta\chi^2$ smaller (larger) than 100 units
is deemed to be accepted (disfavored) at about 90\% C.L.
Thus, a reduction of $\chi^2$ by less than forty units for the BHPS curves has
significance roughly of order one standard deviation. 
The new upper limits on $\langle{x}\rangle_{\rm IC}$ in the CT14 and
CT14HERA2 analyses at the 90\% C.L.:
$\langle{x}\rangle_{\rm IC} \lesssim  0.021 \mbox{ for CT14 BHPS}; ~\langle{x}\rangle_{\rm IC} \lesssim  0.024~\mbox{ for CT14HERA2 BHPS}; 
~\langle{x}\rangle_{\rm IC} \lesssim  0.016~\mbox{for CT14 and CT14HERA2 SEA}. \label{xICconstraints}
$

\subsection{Impact of IC on the electroweak $Z$ and $H$ boson production cross sections at the LHC}

\begin{figure}[tb]
\begin{center}
\includegraphics[width=0.48\textwidth]{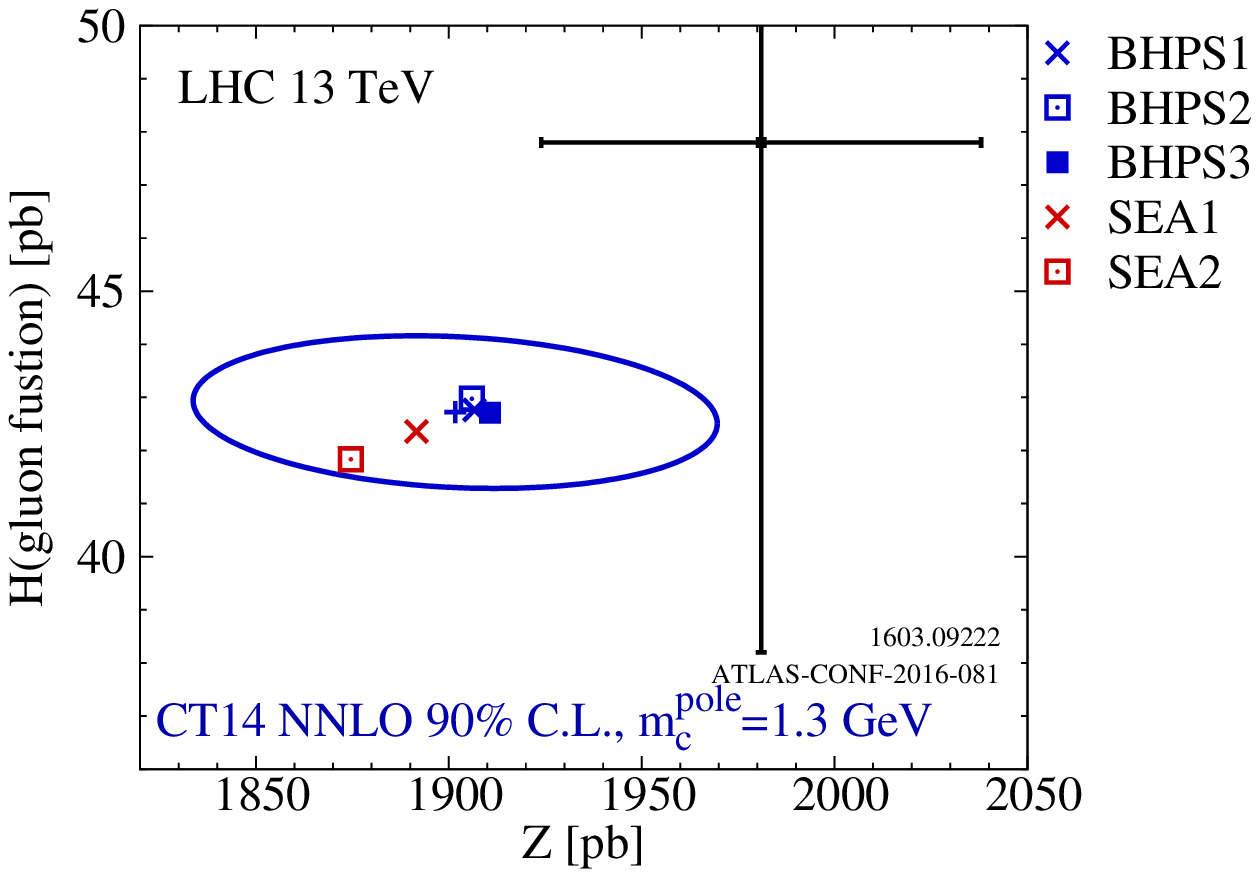}
\includegraphics[width=0.48\textwidth]{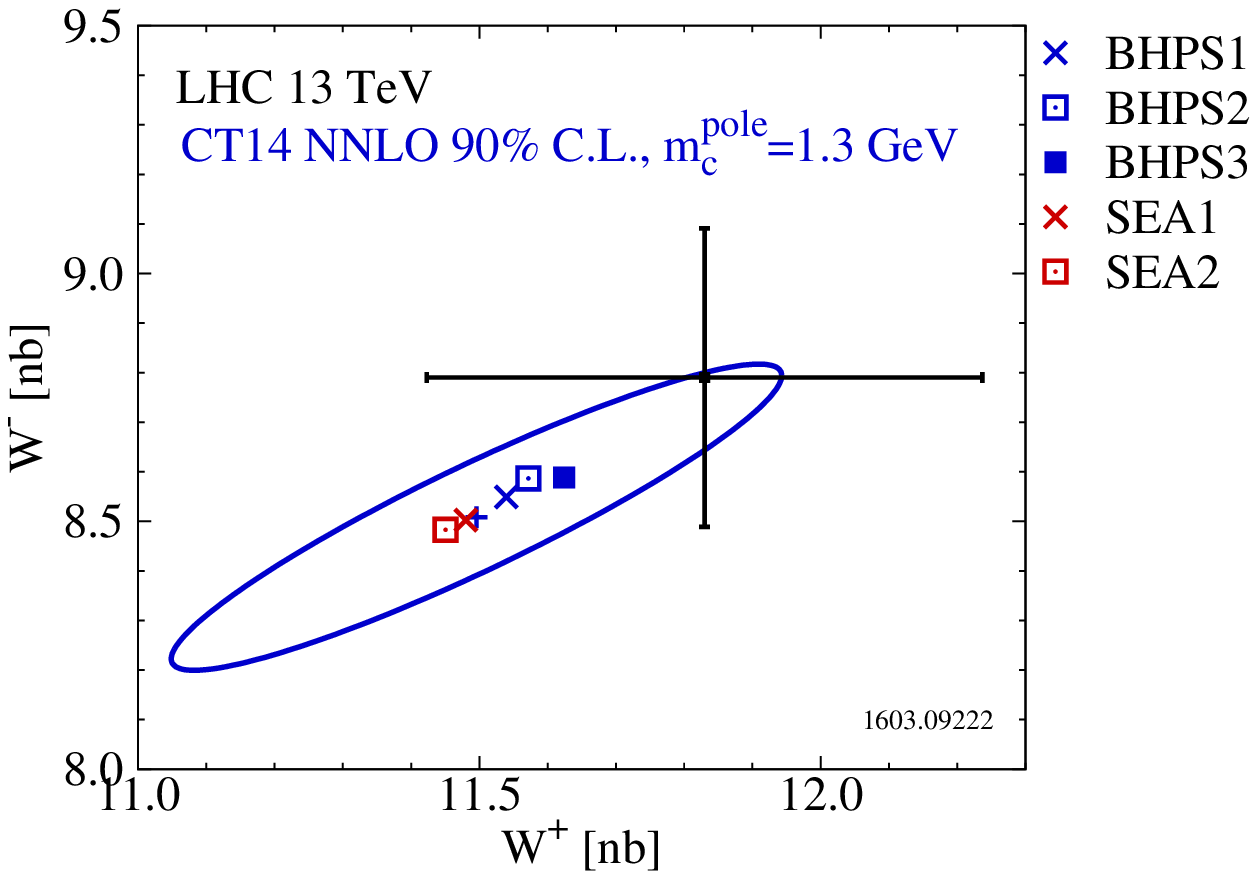}
\end{center}
\caption{CT14 NNLO $H$ (gluon-gluon fusion), $Z$, $W^+$, and $W^-$ production cross sections with an IC PDF component at the LHC $\sqrt{s}$ = 13 TeV, 
with pole mass $m_c^{pole}=1.3$ GeV. The 90\% C.L. uncertainty regions for CT14 at NNLO and experimental points from ATLAS~\cite{Aad:2016naf,ATLAS:2016hru} are also shown. 
\label{fig:ZH-Xsec}}
\end{figure}

Figure~\ref{fig:ZH-Xsec} illustrates predictions of the total cross sections for 
inclusive production of electroweak bosons $W^\pm$, $Z^0$, and $H$ (via gluon-gluon fusion) 
for the BHPS and SEA models at the LHC at a center of mass energy $\sqrt{s}$ of 13 TeV with charm-quark mass $m_c=1.3$ GeV.
The measurements from the ATLAS collaboration~\cite{Aad:2016naf,ATLAS:2016hru} are shown together with 
error ellipses corresponding to the CT14NNLO PDF uncertainties at the 90\% C.L. in order to provide a visual estimate of the impact of the CT14 uncertainties.
%
%
The theoretical prediction for the $W$ and $Z$ inclusive cross sections
(multiplied by branching ratios for the decay into one charged lepton flavor),
is obtained by using the \textsc{Vrap} v0.9 program~\cite{Anastasiou:2003ds,Anastasiou:2003yy} at NNLO in QCD, with the renormalization
and factorization ($\mu_R$ and $\mu_F$) scales set equal
to the invariant mass of the vector boson.
The Higgs boson theoretical cross sections via gluon-gluon fusion at NNLO in QCD
are obtained by using the \textsc{iHixs} v1.3 program~\cite{Anastasiou:2011pi},
in the heavy-quark effective theory (HQET) with finite top quark mass correction,
and with the QCD scales set equal to the invariant mass of the Higgs boson.
The central value predictions for the BHPS and SEA models are all within the CT14 NNLO uncertainties, with BHPS very close to the CT14
nominal fit. The impact of IC on these key LHC observables is mild.

\section {Conclusions}

The CT14IC study explored the possibility of sizeable nonperturbative contribution to charm PDF assuming that factorization for such contributions exists. 
The magnitude of the IC component of the proton has been determined and is found to be consistent with the 
CT14 global QCD analysis of hard scattering data. The new upper bounds on the charm momentum fraction $\langle x\rangle_{\rm{IC}}$ are: $\langle x\rangle_{\rm{IC}} <$ 2\% for BHPS IC, and $\langle x \rangle_{\rm{IC}} <$ 1.6\% for SEA IC, both at 90\% C.L..
As of today, the experimental confirmation of the IC component in the proton is still missing, and data 
from far more sensitive measurements are required to test intrinsic charm scattering contributions at NNLO and beyond.

\Acknowledgements
This work was supported in part by the U.S. Department of Energy under Grant No. DESC0010129;
by the U.S. National Science Foundation under Grant No. PHY-1417326; by
the National Natural Science Foundation of China under the Grant No. 11465018; and by
the Lancaster-Manchester-Sheffield Consortium for Fundamental Physics under STFC Grant
No. ST/L000520/1. The work of M.G. is supported by the U.S. National Science Foundation under Grant No. PHY-1820818.
The work of J.G. is sponsored by Shanghai Pujiang Program.

\end{document}

%% file: econfmacros.tex
\def\beq{\begin{equation}}
\def\eeq#1{\label{#1}\end{equation}}
\def\eeqn{\end{equation}}

\def\beqa{\begin{eqnarray}}
\def\eeqa#1{\label{#1}\end{eqnarray}}
\def\eeqan{\end{eqnarray}}

\let\bar=\overbar

\def\Dslash{\not{\hbox{\kern-4pt $D$}}}
\def\dslash{\not{\hbox{\kern-2pt $\del$}}}

\def\msb{{\bar{\ssstyle M \kern -1pt S}}}

